\documentclass[aps,pre,twocolumn,showpacs,floatfix,preprintnumbers,superscriptaddress,amsmath,amssymb]{revtex4-2}
\usepackage{amsmath}
\usepackage{amssymb}
\usepackage{amsfonts}
\usepackage{graphicx}
\usepackage{epsfig}
\usepackage{epstopdf}
\usepackage{dcolumn}
\usepackage{bm}
\usepackage{array}
\usepackage{mathtools}
\usepackage{url}

\usepackage{subcaption}

\usepackage[colorlinks,bookmarks=false,citecolor=blue,linkcolor=red,urlcolor=blue]{hyperref}

\begin{document}
\title{Evaporative cooling by pulse width modulation (PWM) of optical dipole traps}
\author{S. Sagar Maurya}
\email {Electronic mail: shivsagar.maurya@students.iiserpune.ac.in}
\affiliation{Department of Physics, Indian Institute of Science Education and Research, Pune 411008, India}
\author{Joel M. Sunil}
\affiliation{Department of Physics, Indian Institute of Science Education and Research, Pune 411008, India}
\author{Monu Bhartiya}
\affiliation{Department of Physics, Indian Institute of Science Education and Research, Pune 411008, India}
\author{Pranab Dutta}
\affiliation{Department of Physics, Indian Institute of Science Education and Research, Pune 411008, India}
\author{Jay Mangaonkar}
\affiliation{Department of Physics, Indian Institute of Science Education and Research, Pune 411008, India}
\author{\\Rahul Sawant}
\affiliation{I-Hub Quantum Technology Foundation, Indian Institute of Science Education and Research, Pune 411008, India}
\author{Umakant D. Rapol}
\email {Electronic mail: umakant.rapol@iiserpune.ac.in}
\affiliation{Department of Physics, Indian Institute of Science Education and Research, Pune 411008, India}

\begin{abstract}
We introduce a method for cooling atoms in an optical dipole trap using pulse-width modulation (PWM) technique, without reducing the laser power of the dipole trap. The PWM technique involves digital modulation of the trap at a fixed frequency. The effective time-averaged dipole potential is lowered by adjusting the duty cycle of the modulation, thereby implementing evaporative cooling. We show that, this technique effectively reduces temperature and enhances phase space density. A comparison with the standard method of evaporative cooling has also been made. Apart from the atom loss due to reduction of the effective trapping potential, we observe an additional loss channel originating from the lack of trapping potential during the trap ‘off’ time. This atom loss is observed at different modulation frequencies which are an order of magnitude higher compared to trapping frequency of dipole trap. The PWM technique provides an alternative to traditional evaporative cooling in scenarios where it is preferred that the laser power of the trap should be constant.
\end{abstract}
\maketitle
\section*{Introduction}
Laser cooling and trapping of atoms and ions stands at the forefront of investigations in the discipline of atomic physics, offering tools for scientific exploration \cite{Dutta2023}. A cloud of cold atoms has potential applications across various fields, including atom interferometry \cite{RevModPhys.81.1051,PhysRevLett.117.138501}, atomic clock \cite{PhysRevLett.118.073601,PhysRevA.88.042120}, high-precision spectroscopy \cite{Bothwell2022}, ultra-cold chemistry \cite{doi:10.1080/00268970902724955}, and the investigation of diverse physical scenarios such as  quantum many-body physics \cite{RevModPhys.80.885}, quantum simulator \cite{Bloch2012}, and quantum information processing \cite{García-Ripoll_2005}.\\ 
The first step in laser cooling and trapping begins with the magneto-optical trap (MOT) for generating a cloud of cold atoms \cite{PhysRevLett.59.2631,Dalibard:89,PhysRevLett.69.1741,PhysRevLett.76.2432}. Further reduction of the MOT temperature is limited by photon recoil and light-assisted inelastic collisions. To achieve further reduction in temperature, evaporative cooling is carried out in either a magnetic trap or an optical dipole trap. Far-off-resonance optical dipole traps (FORT) \cite{GRIMM200095} are extensively utilized in experiments for production of cold atoms with higher phase-space density (PSD) and achieving Bose-Einstein condensate (BEC) state. The rapid production of cold atoms with higher PSD has numerous applications, including increasing the cycle rate in precision measurements \cite{doi:10.1126/sciadv.aau7948,PhysRevLett.117.138501} and facilitating rapid data collection in experiments involving cold atoms \cite{PhysRevLett.101.255702,PhysRevE.106.034207}. 
Once a sufficient number of atoms are trapped in the dipole trap, the simplest evaporation process involves gradually reducing the power of the dipole trap \cite{PhysRevLett.87.010404,Saptarishi_Chaudhuri_2007} while optimizing the gain in PSD and minimizing the loss of atoms through three body recombination and background collision \cite{Ketterle1996EvaporativeCO}. However, the rate of evaporation diminishes over time as the reduction in trapping potential slows down the re-thermalization process by elastic collisions \cite{PhysRevA.87.053613}. Therefore, various other techniques, such as modifying the evaporation technique or trap geometry, have been explored to enhance the efficiency of the evaporation \cite{PhysRevA.79.063631}. Evaporative cooling in dipole trap has been implemented through various methods, including: simple crossed dipole traps \cite{PhysRevLett.87.010404,doi:10.1143/JPSJ.81.084004}, hybrid and dimple traps \cite{PhysRevA.79.063631,Mangaonkar_2020,Jacob_2011,PhysRevA.83.013630,Wang:21}, runway evaporation \cite{PhysRevA.79.061406,PhysRevA.78.011604}, single beam microscopic dipole trap \cite{PhysRevA.88.023428} and direct laser cooling in dipole trap \cite{PhysRevLett.122.203202}. Additionally, further cooling techniques utilizing time-averaged potential in dipole trap \cite{Albers2022} and zoom lens traps have been demonstrated \cite{PhysRevA.71.011602}. Also, atoms have been trapped in dipole traps using femtosecond lasers at both macroscopic and microscopic levels \cite{PhysRevA.77.045401,PhysRevA.87.033411}. \\
Here, we present a method for evaporative cooling simply by switching the potential of crossed dipole traps on and off. To systematically reduce the dipole trap potential, we utilize the time-averaged dipole potential with the help of PWM technique. This idea is motivated by technique used in coherent population transfer by femtosecond pulses \cite{PhysRevLett.99.033002} and optimized series of pulses for coherent control\cite{PhysRevLett.131.200801,Xu:22}. By operating at a switching frequency which is much higher than the trap frequencies and adjusting the PWM duty cycle from 100\% to 0\%, we lower the time-averaged dipole potential. This process cools the atoms through evaporation. We also observe how changing the modulation frequencies affects atom loss and temperature after evaporation. Here, we first discuss the theoretical model of our method and the conditions for achieving cooling. Further, the experimental implementation of the PWM technique and the results are discussed.
\section{Time-averaged optical dipole trap}
\begin{figure}
	\centering	
	\includegraphics[width=0.45\textwidth]{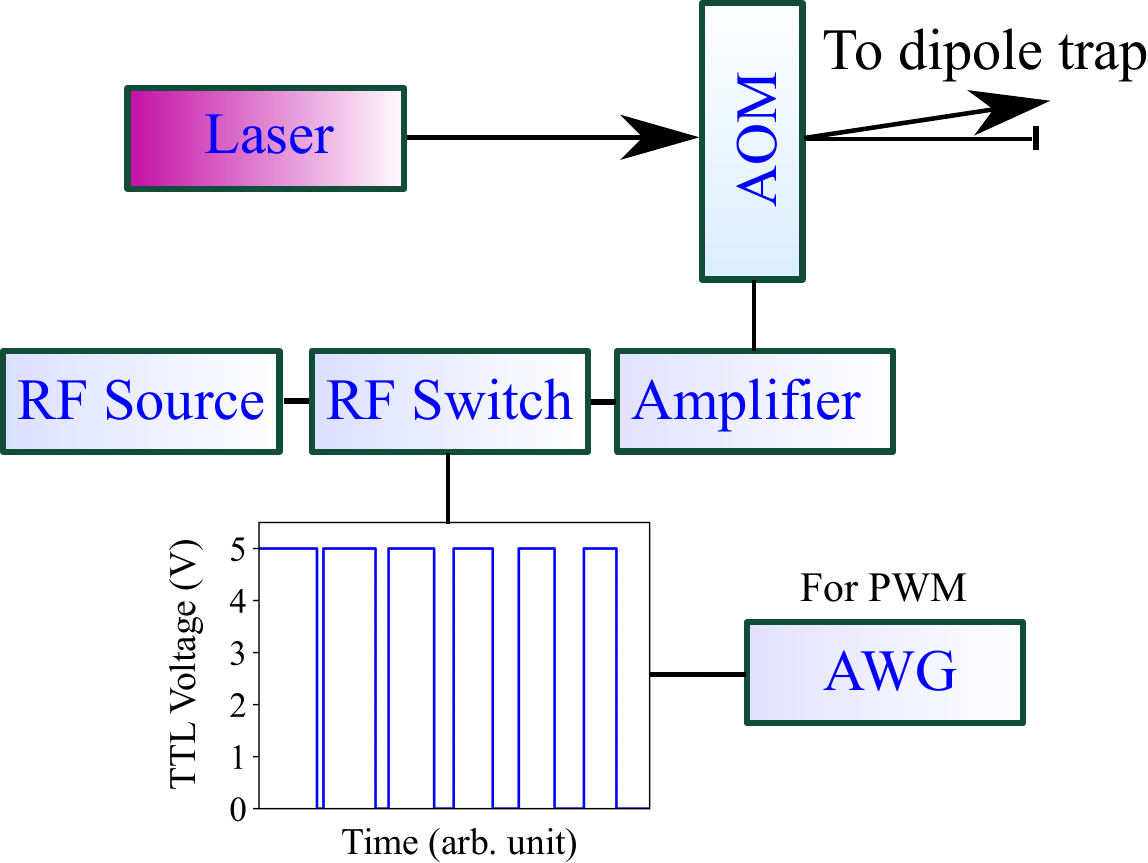}
	\label{fig1}
	\caption{Diagram illustrating the schematic arrangement of the PWM technique utilized for cooling atoms by time-averaged optical dipole trap. }
	\label{Fig1}
\end{figure}

The optical dipole trap potential can be written as \cite{GRIMM200095}:
\begin{equation}
	U_{dipole}(\textbf{r})= -\frac{1}{2\epsilon_{0}c}Re(\alpha)I(\textbf{r})
	\label{eq1}
\end{equation}
where $\alpha$ is the atomic polarizability and $I(\textbf{r})$ is the  electric field intensity.
For the time-averaged dipole potential using PWM technique, we utilize a series of rectangular pulses as shown in Fig. \ref{Fig1}. The rectangular pulses control the switching of laser power on and off through the acousto-optic modulator (AOM). Here, the average intensity of the laser beam varied by changing the duty cycle. This in-turn controls the effective dipole trap potential. Now for a given pulse waveform $f(t)$ with a time-period $T$, the average value of the waveform (in this case, the intensity) is expressed as:
\begin{equation}
	\langle{I}(T)\rangle=  \frac{1}{T}\int_{0}^{T} f(t)\,dt 
	\label{eq2}
\end{equation}

Given that $f(t)$ represents a PWM waveform with a time-period T, when waveform output is high, laser intensity is $I_{\text{max}}$, otherwise it's $I_{\text{min}}$. So for a single cycle, we can write $f(t)$ in terms of laser intensity, like  $I_{\text{max}}$ for $0 < t < D \cdot T$ and $I_{\text{min}}$ for $D \cdot T < t < T$. Here, $D$ can take value between $[0, 1]$ and D = 1 corresponds to $100\,\%$ duty cycle. Consequently, the expression above is transformed to:
\begin{equation}
	\begin{aligned}
		\langle{I}(T)\rangle &=  \frac{1}{T}\left(\int_{0}^{D\cdot T} I_{\text{max}}\,dt + \int_{D\cdot T}^{T} I_{\text{min}}\,dt\right)\\
	&	=  D \cdot I_{\text{max}} +  (1-D)\cdot I_{\text{min}}
	\end{aligned}
	\label{eq3}
\end{equation}

\begin{figure}
	\centering
	\includegraphics[width=0.95\linewidth]{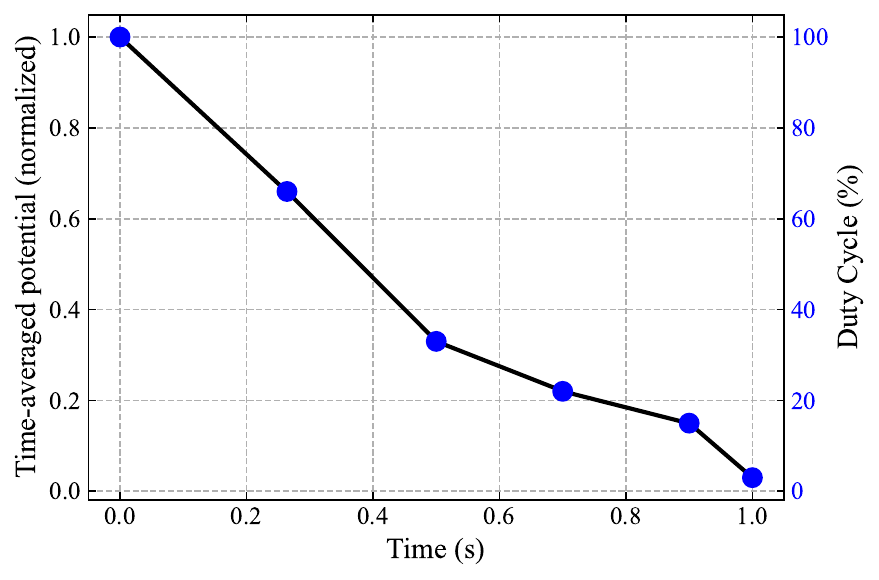} 
	\caption{An optimized time-averaged potential (black solid line) created by numerical integration of PWM waveform. We experimentally generate this potential by changing the duty cycle ( blue markers) linearly in different time-segment.}
	\label{Fig2}
\end{figure}

In this case, where $I_{\text{min}} = 0$, the later expression simplifies significantly to $\langle{I}(T)\rangle = D \cdot I_{\text{max}}$. This implies that the average value of intensity $\langle{I}(T)\rangle$ is determined by $D$ for that cycle and independent of the period.
Thus the time-averaged dipole potential for one cycle with PWM can be written as:
\begin{equation}
	\begin{aligned}
		U_{dipole}(\textbf{r}) &= -\frac{1}{2\epsilon_{0}c}Re(\alpha)\cdot D \cdot I(\textbf{r})\\
	\end{aligned}
	\label{eq4}
\end{equation}
For generating a decreasing time-averaged dipole potential for evaporative cooling, a linear ramp down in $D$ is applied across different time segments. An example of the time-averaged potential is shown in Fig. \ref{Fig2}, where the period of the PWM is $1 \, \mu$s, and the total evaporation time is 1 s. For a period of $1 \, \mu$s, the corresponding modulation frequency is 1 MHz, and we will refer to the modulation frequency throughout this paper. The solid line in Fig. \ref{Fig2} is plotted by integrating each cycle over a total of $1 \times 10^6$ cycles, which represents the time-averaged potential during evaporation. To achieve this experimentally, we linearly ramp $D$ in five segments, starting from D=1 and ending at D=0 (completely off). It's shown with blue markers in Fig. \ref{Fig2}. The time-averaged dipole potential (also shown in Fig. \ref{Fig2}) represents the optimized evaporation sequence for our experimental setup. The sequence is optimized by monitoring the temperature and loss of a atoms. This optimized sequence is used with different modulation frequencies throughout the paper.

\subsection{Condition for cooling with PWM technique }
As the dipole trap laser power is kept constant in this process, modulation frequency plays a major role in evaporative cooling. The modulation frequency is determined by considering the beam waist of the dipole trap and trap depth. If the modulation frequency is too low, the atoms could drift too far and be lost. For effective evaporative cooling, atom loss through elastic collisions should dominate rather than atom loss due to their movement caused by mean temperature. So to calculate the lower limit for modulation frequency by considering the atom's velocity ($v_{atom}$) and beam waist ($2\omega_{0}$) as discussed here \cite{epub93330}, we set a lower bound of:
\begin{equation}
	\begin{aligned}
	f_{mod}\gg \frac{1}{2\omega_{0}}\sqrt{\frac{2k_{B}.T_{trap}}{m}}
    \end{aligned}	
	\label{eq5}
\end{equation}
From the above calculation, the minimum modulation frequency($f_{mod}$) for our setup is 3.5 kHz. So, for effective evaporation through time-averaged potential, the modulation frequency should be higher than by at least 10-100 times \cite{PhysRevLett.99.083001}. To validate this, we measure the loss of the atoms with different modulation frequency by holding the atoms in trap. We scan the modulation frequency from 200kHz to 1 MHz, find the minimum loss corresponding to the 1 MHz.

\section{Experimental Details}
The experimental sequence begins with loading $\sim 2 \times 10^{6}$ atoms of $^{87}\text{Rb}$ in a crossed dipole trap at a temperature of $120~\mu\text{K}$ from a standard 3D-magneto-optical trap. The wavelength of the crossed dipole trap beam is 1064 nm and beam waist is $100~\mu\text{m}$  at focus point with a power of $5$ Watt in each beam. After loading the atoms in dipole trap, we switch off the MOT loading and the magnetic-field gradient ($22~\text{G/cm}$) is kept on during the evaporation process to support the atoms against the gravity. We used a 120 MHz and 110 MHz acousto-optic modulator (AOM) to drive the dipole beams and PWM is applied through these AOMs as shown in Fig. \ref{Fig1}. More details about our experimental setup is described here \cite{PhysRevA.110.053307}.
 
For the evaporative cooling, PWM pulse was continuously applied throughout the entire sequence of evaporation, spanning from the initial step to the final step. We use a arbitrary waveform generator (AWG:Spectrum Instrumentation-M4i.6631-x8), externally programmed for generating and shaping the PWM pulse for duty cycle ranging from $100\,\%$ to close to $0\, \%$. Further, the output signal from the AWG was connected to a fast RF switch which controls the switching of AOMs for time-averaged dipole potential. In the final stage, as we gradually reduce the duty cycle, we encounter limitations imposed by the rise and fall time of the AOMs, which is approximately 100 ns. The evaporation efficiency is optimized by measuring the temperature ($T$) by time-of-flight method. The number of atoms ($N$) and the atomic cloud expansion is determined by absorption imaging. 
\section{Results} 
The first step for PWM cooling is to determine the modulation frequency at which the losses discussed in the previous section are minimized. This frequency is determined by observing the atom loss under modulation of the dipole trap for one second of hold time. The modulation frequency is varied from $200$ kHz to $1$ MHz with $50 \%$ duty cycle. We observe losses at at $200$ kHz even when it is orders of magnitude higher compared to the limiting frequency of $3.5$ kHz. As shown in Fig. \ref{Fig2}, the loss decreases as the modulation frequency is increased. We gain almost $11$ \% atoms per $100$ kHz increment in modulation frequency. 
\begin{figure}
	\centering
	\includegraphics[width=\linewidth]{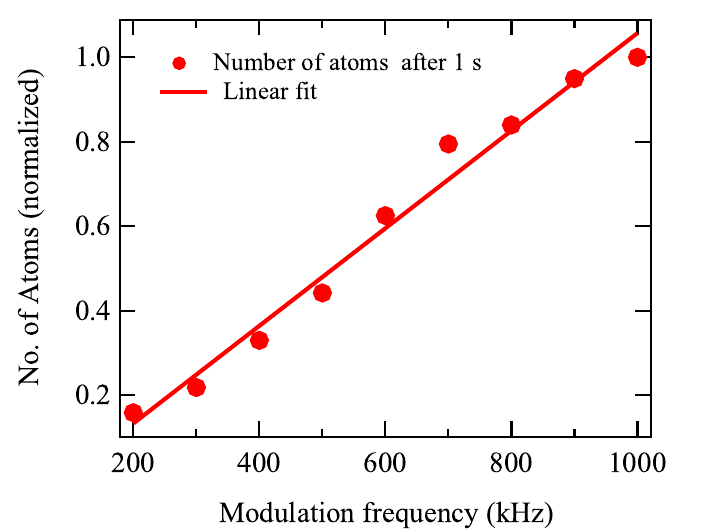} 
	\caption{No. of atoms after 1 s hold time with different modulation frequency and fixed 50\% duty cycle. Here, markers are experimental data and solid line is linear fit.}
	\label{Fig3}
\end{figure}
At $1$ MHz modulation  we observe that the atom number loss is very close to the case of not modulating the trap. To ensure that the loss dynamics are similar to the case of not modulating the trap, we do further experiments. 
\begin{figure}
	\centering
	\includegraphics[width=\linewidth]{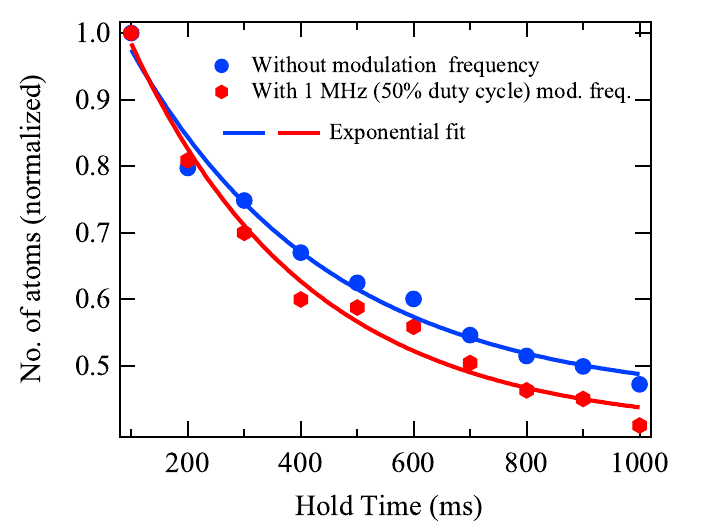} 
	\caption{Comparison of the loss of the atoms with and without modulation at fixed power. Here, markers are experimental data and solid lines are exponential fit.}
	\label{Fig4}
\end{figure}

\begin{figure}
	\centering
	\includegraphics[width=\linewidth]{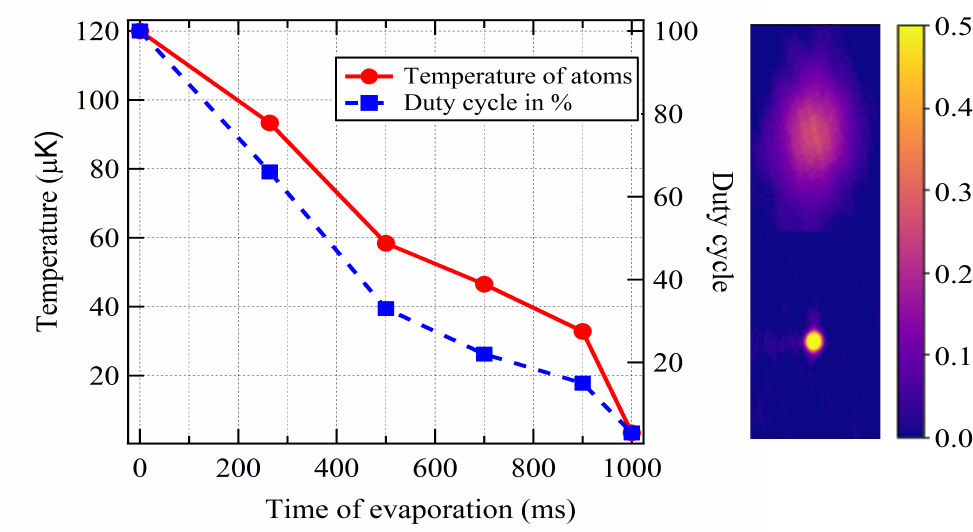} 
	\caption{Evaporative cooling is shown with a decreasing duty cycle over time in blue, corresponding to a decrease in temperature in red. A time-of-flight image of cold atoms on the right, showing the state before evaporation and after 1 second of evaporation by PWM.}
	\label{Fig5}
\end{figure}

We compared the atom loss rate with and without modulation. For the case of no modulation in the trap, we reduced the dipole trap power by 50\% to match the time-averaged potential with modulation. As seen in Fig. \ref{Fig4}, for the case of no modulation, the loss rate was slightly lower compared to modulating it at 1 MHz, when the average power in both cases is the same. The atoms loss rate is comparable in both case with time constant of $350$ ms for without modulation and $313$ ms for with modulation. In our setup, density dependent losses dominates in both case. We start evaporative cooling by creating time-averaged potential. The evaporation ramp is divided into five steps. At each step the average trapping potential is decreased by changing the duty cycle in a linear fashion. It's shown in Fig. \ref{Fig5} with blue dashed line.
Each evaporation step is optimized manually by observing the reduction in temperature, which is also plotted in Fig. \ref{Fig5} by red line. The corresponding time-averaged potential by integration is shown in Fig. \ref{Fig2}. The lowest temperature that we achieve by this manual optimization is about $3\, \mathrm{\mu K}$. Here, machine learning techniques \cite{Biswas2023,Wigley2016} for time-averaged potential could also be implemented for reducing the temperature.  
\begin{figure}[h]
	\centering
	\includegraphics[width=1\linewidth]{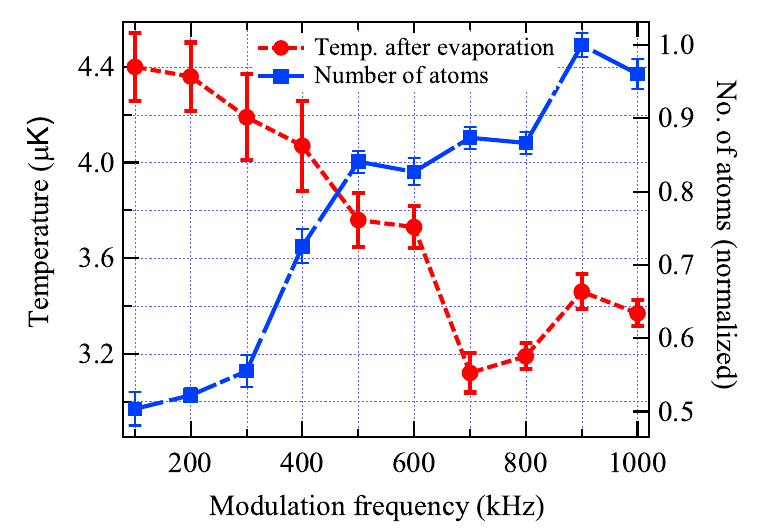} 
	\caption{Temperature in red after full evaporation at different modulation frequency , no. of atoms in blue is shown at right axis.}
	\label{Fig6}
\end{figure}
Finally, we measure the final temperature and no. of atoms after full evaporation, following same evaporation sequence as in Fig. \ref{Fig5} by changing the modulation frequency from 100 kHz to 1000 kHz.
The data is shown in Fig. \ref{Fig6}, each point is averaged with 10 data set. We observe a gradual decrease in temperature by frequency with a minima at 700 kHz. The increase in modulation frequency reduces the temperature as well as enhance the atoms number, for the no. of atoms, a connection with Fig. \ref{Fig3} could be build. Here, We get minima in temperature at 700 kHz because more smoother time-average potential, which get compromised with AOM response time at higher frequency at the end of the sequence. At lower frequencies, atom loss may have compromised the cooling. We also measure the temperature after standard evaporation, which we get $3\, \mathrm{\mu K}$, following the same dipole potential reduction as the time-averaged potential over $1$ s. Both methods yield similar results within this time constraint of $1$ s.
\begin{figure}[h]
	\centering
	\includegraphics[width=0.85\linewidth]{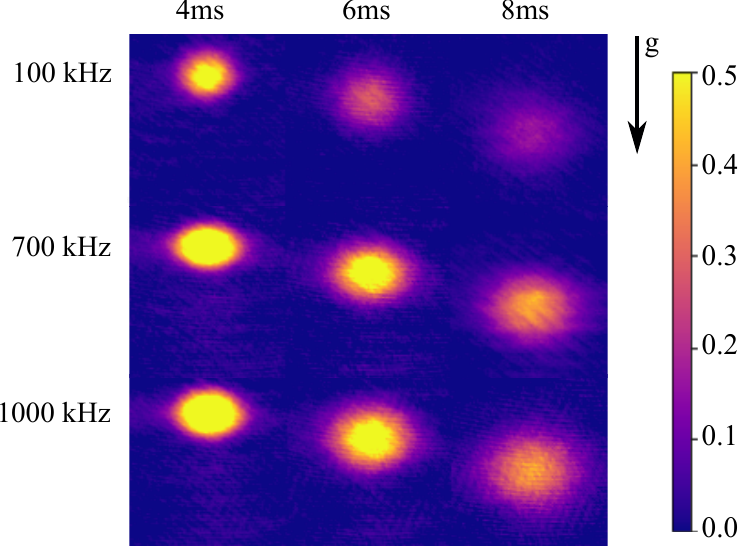} 
	\caption{(a) Time of flight images of atomic cloud at different modulation frequency (b) Rate of expansion of cloud width with different time-of-flight for above three cases.}
	\label{Fig7}
\end{figure}
For measuring the temperature of atoms as shown in Fig. \ref{Fig6} after evaporation, we measure the expansion rate of cloud width by providing different time-of-flight. The relation between cloud width and temperature are given by \cite{PhysRevA.101.013420}:
\begin{equation}
	\begin{aligned}
	\sigma(t) = \sqrt{\sigma_0^2 + \frac{k_B T t^2}{m}}
    \end{aligned}
    \label{eq6}
\end{equation}
In Fig. \ref{Fig7}, we have shown some example images of atomic cloud for (100, 700 and 1000) kHz modulation frequency for (4, 6 and 8) ms  time-of-flight. A clear difference in temperature as well as atoms loss could be seen. Further, with this evaporation technique, we achieve an approximately four order of magnitude increase in phase space density, reaching value close to $10^{-2}$ for 700 kHz. It's in same proportion for other modulation frequencies, corresponding to their temperatures and atom numbers.
\section{Conclusion}
In this experiment, we introduced and demonstrated a method for cooling atoms in optical dipole traps without reducing the laser power of the dipole traps. Our approach involves creating a time-averaged dipole potential by rapidly switching the optical dipole traps on and off and adjusting the duty cycle by the PWM technique at a fixed frequency via an RF switch. By carefully optimizing the duty cycle, we successfully lowered and shaped the time-averaged dipole potential, facilitating atom cooling through evaporative processes. We reduce the temperature of cold atoms from 120 $\mathrm{\mu K}$ to 3 $\mathrm{\mu K}$ in $1$ s, keeping over 10 \% of the initial atoms. This method is as efficient as standard evaporation in terms of both temperature reduction and atom retention.

In conclusion, PWM-based evaporative cooling method presents a robust, efficient, and easily implementable approach for atom cooling in optical dipole traps. This technique's simplicity and digital control simplify cooling individual dipole traps.  The lower temperature of atoms in dipole trap enhance photo-association efficiency for neutral atom arrays in quantum computation \cite{Wang2020,PhysRevLett.115.073003}. PWM technique can also be implemented in effective evaporative cooling in  micro-gravity as in this recent experiment \cite{PhysRevA.101.013634}. It can perform better due to the absence of gravitational pull when the dipole beams are off. Further, eliminating light shift in precision spectroscopy \cite{Hutzler_2017} is also possible by this technique .

\section{Acknowledgments}
\addcontentsline{toc}{section}{Acknowledgments}
S.S.M. and J.M.S acknowledge financial support through a research
fellowship from Council of Scientific and Industrial Research,
Government of India. The authors acknowledge the fruitful discussions with Korak Biswas and Kushal Patel. All the authors are grateful to the National Mission on Interdisciplinary Cyber Physical Systems for funding from the DST, Government of India through the I-HUB Quantum Technology Foundation, IISER Pune.

\bibliography{Ref}
\end{document}